\documentclass[a4paper,11pt]{article}
\pdfoutput=1
\usepackage{jinstpub}

\title{Measurement of the Ionization Yield of Neutron-Induced Proton Recoils in Tetramethylsilane}

\author[a]{S.X.~Wu,\note{Corresponding author.}}
\author[a]{B.G.~Lenardo,}
\author[a]{G.~Gratta}

\affiliation[a]{Physics Department, Stanford University, 382 Via Pueblo Mall, Stanford, CA 94305, USA}
\emailAdd{sxwu@stanford.edu}

\abstract{
We report on a low energy measurement of the ionization yield in a Tetramethylsilane Time Projection Chamber (TPC) using 2.8 MeV neutrons from a deuterium-deuterium neutron generator. The proton recoil charge yield is measured at four different electric fields, finding a dependence that is well described by the Thomas-Imel model. By comparing the proton recoil yield to that obtained from $\gamma$-ray calibrations, a quenching factor is obtained for each electric field. These results demonstrate the feasibility of using room temperature organic ionisation detectors to detect MeV-scale neutrons in the proton-recoil channel.}

\begin{document}

\maketitle

\section{Introduction} 

Liquid-phase ionization chambers are widely-used for the detection of radiation, due to their characteristics of fine segmentation, good energy resolution, and scalability to large detectors. Often, the medium is a cryogenic noble liquid such as liquid Argon~\cite{ATLAS_calo}, Xenon~\cite{meg_exp} or Krypton~\cite{NA62} because of their demonstrated ability to achieve long electron lifetimes, access to both ionization and scintillation, and good ionization yield. However, the use of cryogenic systems greatly complicates the design and operation of these detectors. Detectors based on room-temperature, organic liquids have the potential to combine the advatages of liquid ionization chambers with a simpler design. In addition, molecular liquids offer a wider variety of atomic composition of the target material, offering the possibility to select liquid properties to suit specific applications.

Organic liquids were explored in the 1980's and 1990's for calorimetry in high energy physics, driven primarily by the upgrade of the UA1 calorimeter~\cite{UA1_TMP_Calo}. At high energies, the Tetramethylpentane (TMP) - Uranium calorimeter offered fine granularity (useful for high multiplicity final states), full event containment, good energy resolution, uniformity, hermeticity, and equal response for electrons and hadronic showers~\cite{UA1_TMP_Calo}.
As a part of this development program, the ionization response of organic liquids to protons and other highly-ionizing particles was studied for incident energies above 8~MeV~\cite{DUHM1989565}. In addition, a dependence of the ionization signal on the angle between the track and the electric field was measured in Tetramethylsilane (TMS) for high-energy, heavily-ionising particles~\cite{DUHM1989565, ENGLER1992479}, suggesting the existence of columnar recombination in organic liquids.

At lower energies, the availability of free protons in an organic liquid could enable new applications in the detection of fast neutrons via elastic-scattering-induced proton recoils, or the detection of antineutrinos via the inverse beta decay process (for which fast neutron scattering is often an important background). Hydrogen-rich target materials could also offer a new tool to search for nuclear recoils induced by low-mass ($\mathcal{O}(1~ \mathrm{GeV})$) Weakly Interacting Massive Particles (WIMPs), thanks to the light nuclear mass and thus higher kinetic energy transferred. These applications would be further enhanced by the columnar recombination, which could provide directional information in the recoil signal. It is therefore of considerable interest to characterize the ionization response of TMS to proton recoils at energies relevant for these types of experiments, i.e. at a few MeV and below.
 
This work reports a new measurement of the ionization response and the ``quenching factor'' of TMS for proton recoils induced by the scattering of 2.8~MeV neutrons.  
The measurement is obtained with the detector described in~\cite{TMS_paper}, using monoenergetic neutrons from a deuterium-deuterium (D-D) fusion generator. The ionization yield is measured by fitting the endpoint of the ionization spectrum to determine the signal size at the maximal energy deposition. This is scaled to units of absolute charge by comparing to previous measurements, and a proton recoil quenching factor is obtained. The measurements are repeated at four different electric fields to study the charge recombination inside TMS.

\section{Experimental Setup}

\subsection{TMS Time Projection Chamber}

The detector used in this work is a multi-wire time projection chamber (TPC), described in detail in Ref.~\cite{TMS_paper}. It consists of a $\sim 100\times 100$~mm$^2$ readout surface and a 15.5\,mm drift length, housed inside of a 150.4\,mm-diameter cylindrical ultra-high vacuum vessel. During operation, the vessel is filled with $\sim$1 liter of liquid TMS maintained at 22$\pm$0.5$^{\circ}$C. One end of the drift region is defined by a stainless steel cathode plate, while the other is defined by two orthogonal wire planes which sense the ionization signal. A uniform electric field is generated across the drift volume by applying negative high voltage (HV) to the cathode while maintaining the first and second wire planes at ground and positive potential, respectively. The bias between wire planes is set so that charge passes freely through the first and is fully collected by the second. Each plane consists of ten wire triplets on a 9~mm pitch. The wires in each triplet are electrically connected (with a 3~mm pitch) providing better field uniformity with a smaller readout channel count.  The central eight triplets are instrumented with charge-sensitive preamplifiers identical to those described in Ref.~\cite{Jewell_2018} (and based on the design described in Ref.~\cite{Fabris_1999}), while the two outermost triplets are biased as guard wires but not read out. The first plane, ``Y-wires'', records the induction signal from charges drifting past, while the second, ``X-wires'', collects and measures the ionization. The energy of an event inside the TPC is reconstructed by the total charge collected, while position information is reconstructed from the combination of X and Y signals measured above noise. The detector also uses a 5-inch diameter hemispherical photomultiplier tube (PMT)\footnote{ET Enterprises, model 9372B} mounted onto a viewport, to detect the Cherenkov light signals produced in the TMS. In the present work, the recoiling protons are well below the Cherenkov emission threshold, and the PMT is used as a veto to reduce $\gamma$-ray backgrounds, as described in Section~\ref{subsec:reconstruction_and_selection}.

\subsection{Neutron generator and shielding}
\label{subsec:shielding}

\begin{figure}[h!]
    \centering
    \includegraphics[width=0.85\columnwidth]{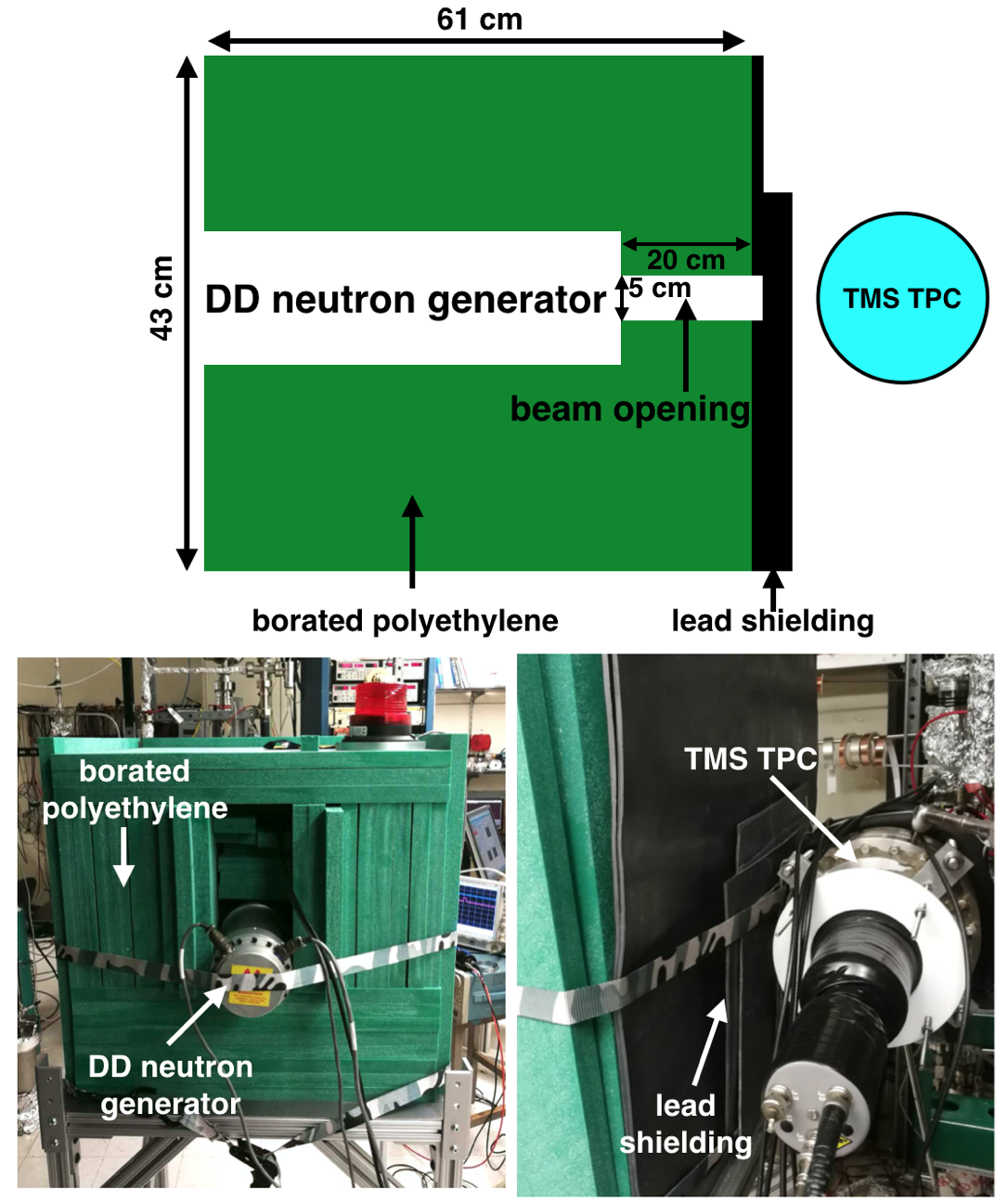}
    \caption{Experimental setup. Top: schematic of the setup; bottom left and right: back and front view of the neutron generator, borated polyethylene collimator, lead shielding, and TPC with Cherenkov light PMT.}
    \label{fig:ExpSetup}
\end{figure}

The overall experimental setup is shown in Fig.~\ref{fig:ExpSetup}. Neutrons are produced by a ThermoFisher Scientific MP320 D-D generator via the D(d,n)$^{3}$He fusion reaction. The generator head is placed inside a custom-designed borated polyethylene (BPE) shield and collimator with a ~5~cm diameter beam aperture meant to select neutrons directed towards the TPC. Neutrons emitted off-axis are thermalized in the BPE via (n,p) elastic scattering and then captured on $^{10}$B. The accelerating potential was maintained at 80~kV during data taking, resulting in a production rate of $\sim$10$^6$~n/s with kinetic energy of 2.8~MeV in the forward direction~\cite{LISKIEN1973569}. The generator was operated in pulsed mode, with a pulse duration of 10~$\mu$s and a duty cycle of 15\%, with $\sim0.05$~n/pulse reaching the detector.

The $^{10}$B capture reaction emits $\gamma$s at a relatively low energy of 0.48~MeV. In order to reduce the background rate from these $\gamma$s, lead shielding generally covers the face of the BPE collimator towards the detector. This lead shield is made out of 3.2~mm thick plates for a total of 25.6~mm in most places.   Because the lead plate is recycled from a previous experiment, the shield is thinner at the collimator opening and at the top, as schematically shown in Figure~\ref{fig:ExpSetup}.

\subsection{Data acquisition}

The output signals from the 8+8 preamplifiers instrumenting the X and Y wire grids are split into two branches for digitization and triggering. The first branch, along with signals from the PMT and the beam gate from the neutron generator, are digitized at 62.5~MS/s with two Struck 3316 16-bit digitizers. The data acquisition window is set to 128~$\mu$s (8000 samples) with 32~$\mu$s (2000 samples) before the trigger, to be used for baseline calculation. The second branch of signals is sent to the trigger logic, which requires an ionization signal in the TPC in coincidence with the beam gate. Since the noise depends somewhat on the channel and, in particular, on the housing containing groups of four channels of preamplifiers~\cite{TMS_paper}, the trigger only uses the three lowest-noise channels (X2, X3, X4)  which are fed to a shaping amplifier~\footnote{Ortec 572A} with a shaping time of 1~$\mu$s. The unipolar output is then fed into an oscilloscope~\footnote{Agilent DSO7014A}, used as a discriminator. A threshold of twice of the root-mean-square (RMS) noise of the signal after the shaping amplifier is used to trigger the oscilloscope. 
The beam gate is formed as a 15~$\mu$s pulse, accounting for the 10~$\mu$s D-D generator pulse plus the maximum drift time in the TMS ($\sim$3~$\mu$s at 8kV cathode voltage), the 1~$\mu$s shaping time, and 1$\mu$s of extra allowance.  The Struck digitizers are triggered by the coincidence between the OR of the signals in X2, X3, X4 and the beam gate.  

\begin{figure}[h!]
    \centering
    \includegraphics[width=0.8\columnwidth]{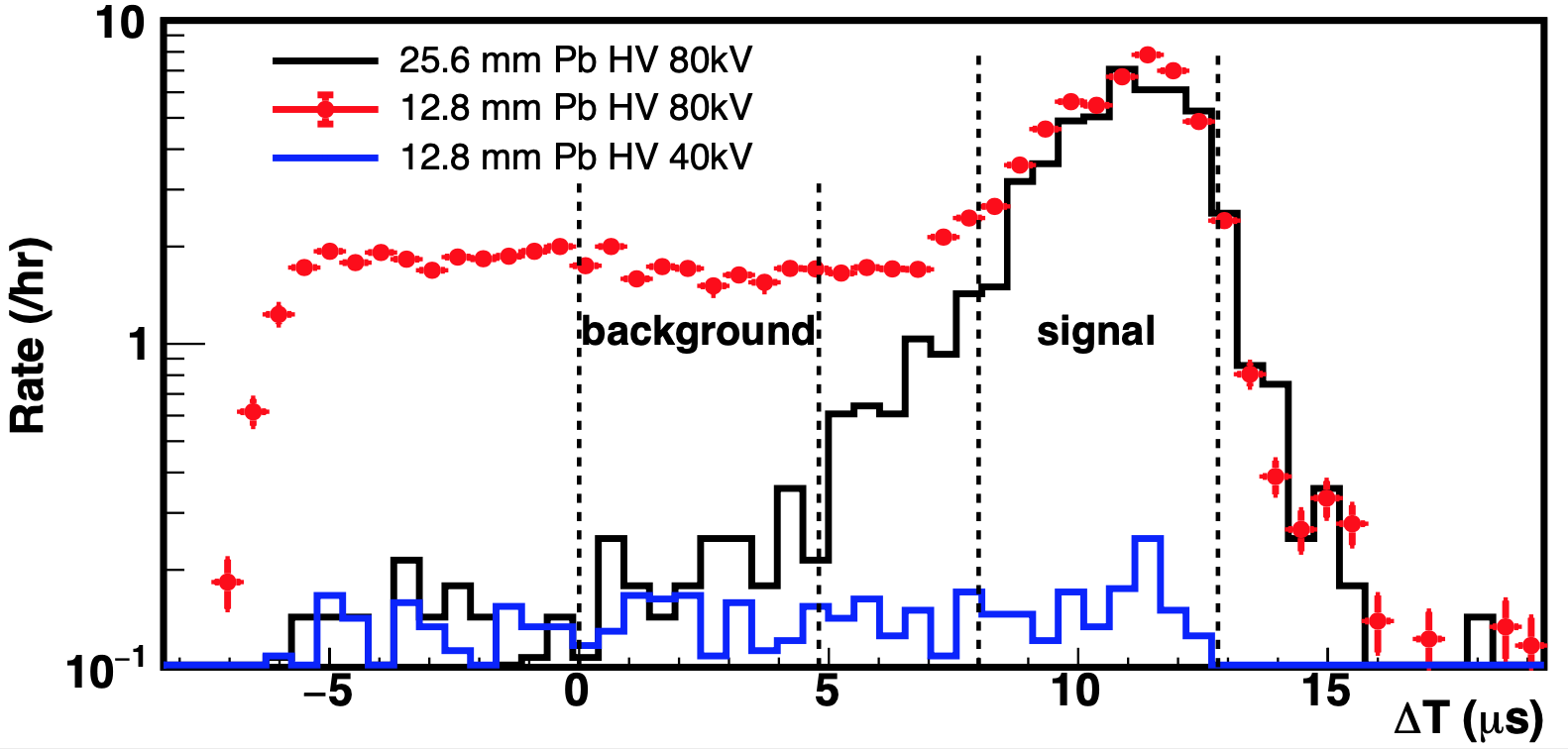}
    \includegraphics[width=0.8\columnwidth]{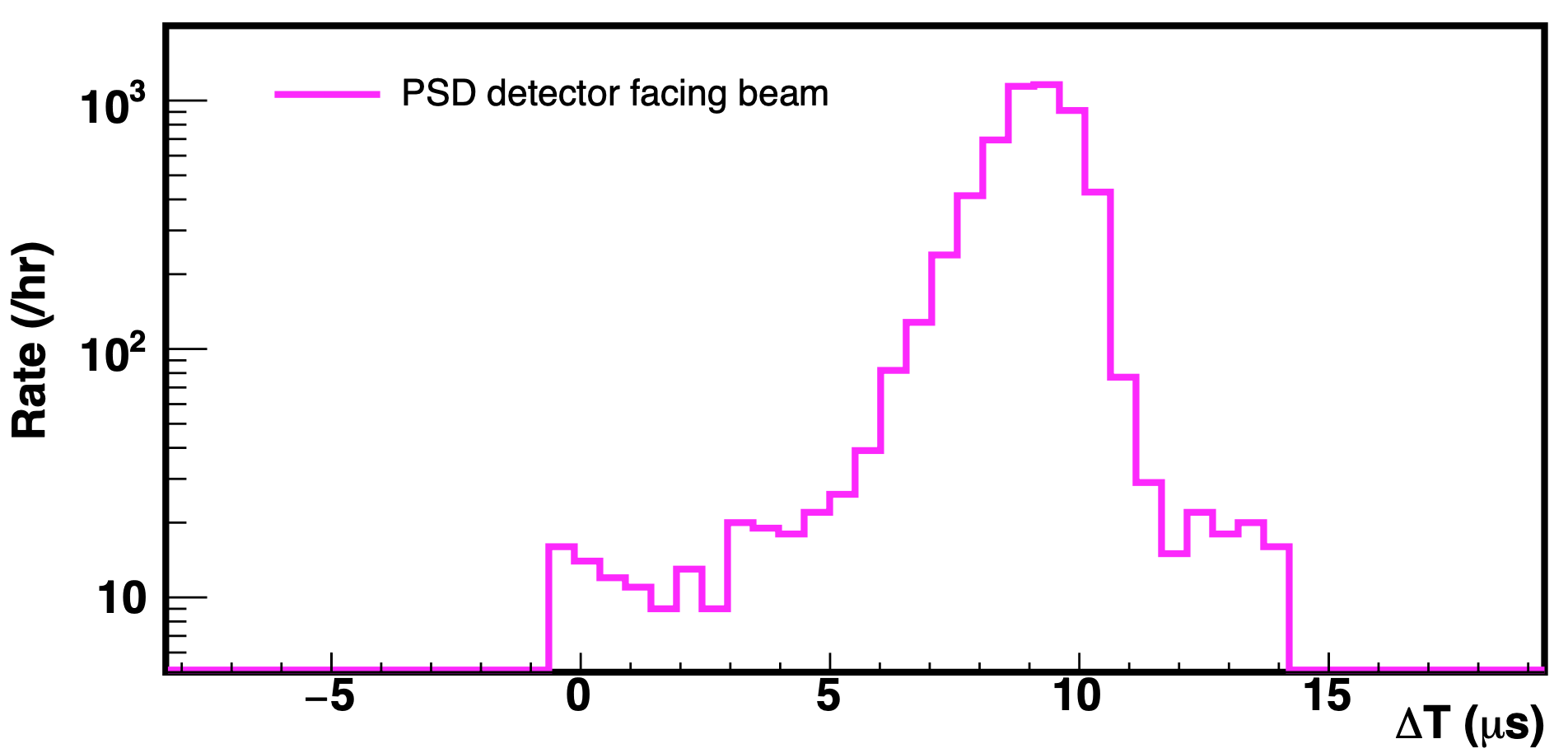}
     \caption{Time difference between the signals in the TMS detector (top) or the PSD scintillator (bottom) and the beginning of the neutron beam gate. The lead shielding thickness and the HV setting of the neutron generator are denoted in the legend (top).  }
 \label{fig:timing}
\end{figure}

Fig.~\ref{fig:timing} (top) shows the distribution of the time difference between the ionization signal and the start of the beam gate for different sets of parameters. A clear peak is observed between 8--13~$\mu$s after the trigger in the nominal data taking configuration (labeled ``25.6~mm Pb, HV 80~kV"). This is confirmed to be dominated by neutron scattering in the TMS by replacing the TPC with a scintillator capable of pulse shape discrimination (PSD)~\footnote{EJ-276, Eljen technology}. An event time structure consistent with the one recorded in TMS occurs in the PSD scintillator, as shown  in Figure~\ref{fig:timing} (bottom).   The timing distribution of the TPC signals is shifted by $\sim$3$~\mu$s and broadened compared with the PSD detector due to the drift time and broader timing resolution in the TPC.  The adequacy of lead shielding thickness is empirically verified by a run with only 12.8~mm of lead, also shown in Fig.~\ref{fig:timing} (top), where a substantial number of signals recorded at earlier time is evident, owing to the faster propagation times of $\gamma$s produced inside the BPE.  Also shown in Fig.~\ref{fig:timing} (top) is the time spectrum for a run done with the accelerating potential of the D-D generator set to 40~kV, for which the neutron production rate is expected to be negligible.  The conclusion of this preliminary study is that, in the nominal data taking configuration, events beyond $\sim$5~$\mu$s after the trigger are indeed dominated by neutron interactions in the TMS, with the regions selected to measure the neutron signal and the background marked in Fig.~\ref{fig:timing} (top).

\begin{figure}[h!]
    \centering
    \includegraphics[width=0.85\columnwidth]{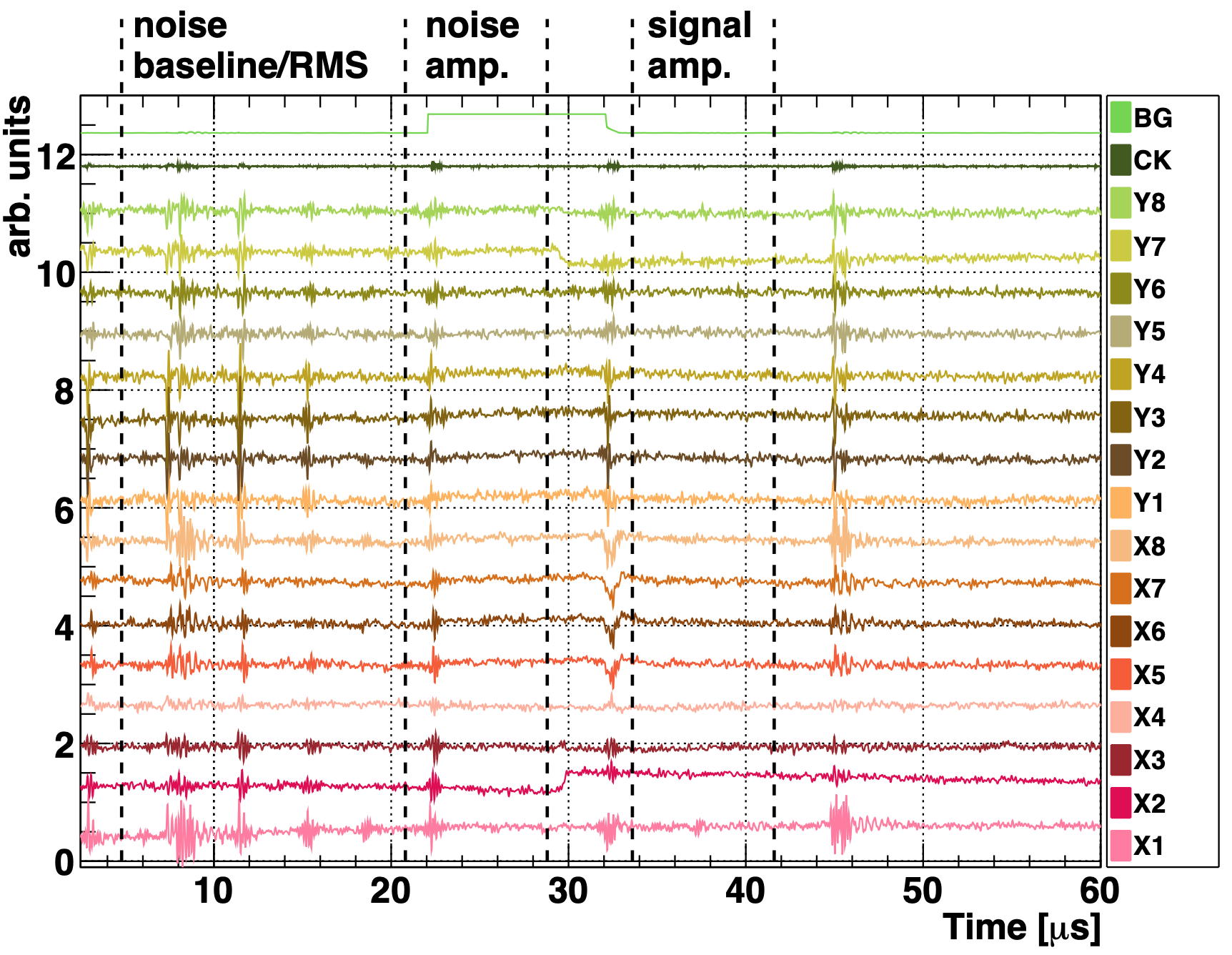}
    \caption{Example of digitized signals from a proton recoil candidate event in TMS. From the top are the beam gate denoted as BG and the Cherenkov PMT signals denoted as CK, followed by 16 waveforms from the Y and X wires. 
    The vertical scale is arbitrary and different for the ionization signals, PMT and the logic beam gate. The time periods used for the noise baseline/RMS, noise amplitude and signal amplitude calculations separated by vertical dashed lines and identified at the top. The periodic noise patterns on the wire signals are related to the operation of the neutron generator and do not appreciably affect the signal amplitude calculation because they are of short duration and average to zero.  A charge collection signal with positive polarity is visible in channel X2 with a corresponding induction signal with negative polarity is in channel Y7. Note that there is no PMT pulse in this event.}
    \label{fig:eventexample}
\end{figure}

In total, data is recorded at four different cathode voltages: 6, 7, 8 and 9~kV, corresponding to drift fields of 3.9, 4.5, 5.2 and 5.8~kV/cm in the TPC volume. An example of digitized waveforms from a neutron-induced event is shown in Figure~\ref{fig:eventexample}.

\section{MC simulation and data analysis}

\subsection{Geant 4 simulation}

A simulation of the setup using Geant4~\cite{Geant4_1} is performed to estimate neutron capture backgrounds and model the proton recoil energy spectrum. The simulated geometry includes a detailed model of the internal structure of the TPC, as well as a full model of the BPE and lead shielding described in Section~\ref{subsec:shielding}. The D-D generator is modeled by producing 2.8~MeV neutrons at a point source in the plane of the target.  Neutrons are then propagated in the beam channel towards the TMS detector. To avoid unnecessary computation time, the initial neutron directions are confined to a cone of half-angle 10$^{\circ}$ directed along the beam axis. This substantially exceeds the collimation aperture, so that a representative sample of neutron captures in the BPE are simulated for background estimation purposes. The active TMS volume is subdivided into 9~mm-pitch subvolumes to model the detection of charge from individual X channels.

\begin{figure}[h!]
    \centering
    \includegraphics[width=0.9\columnwidth]{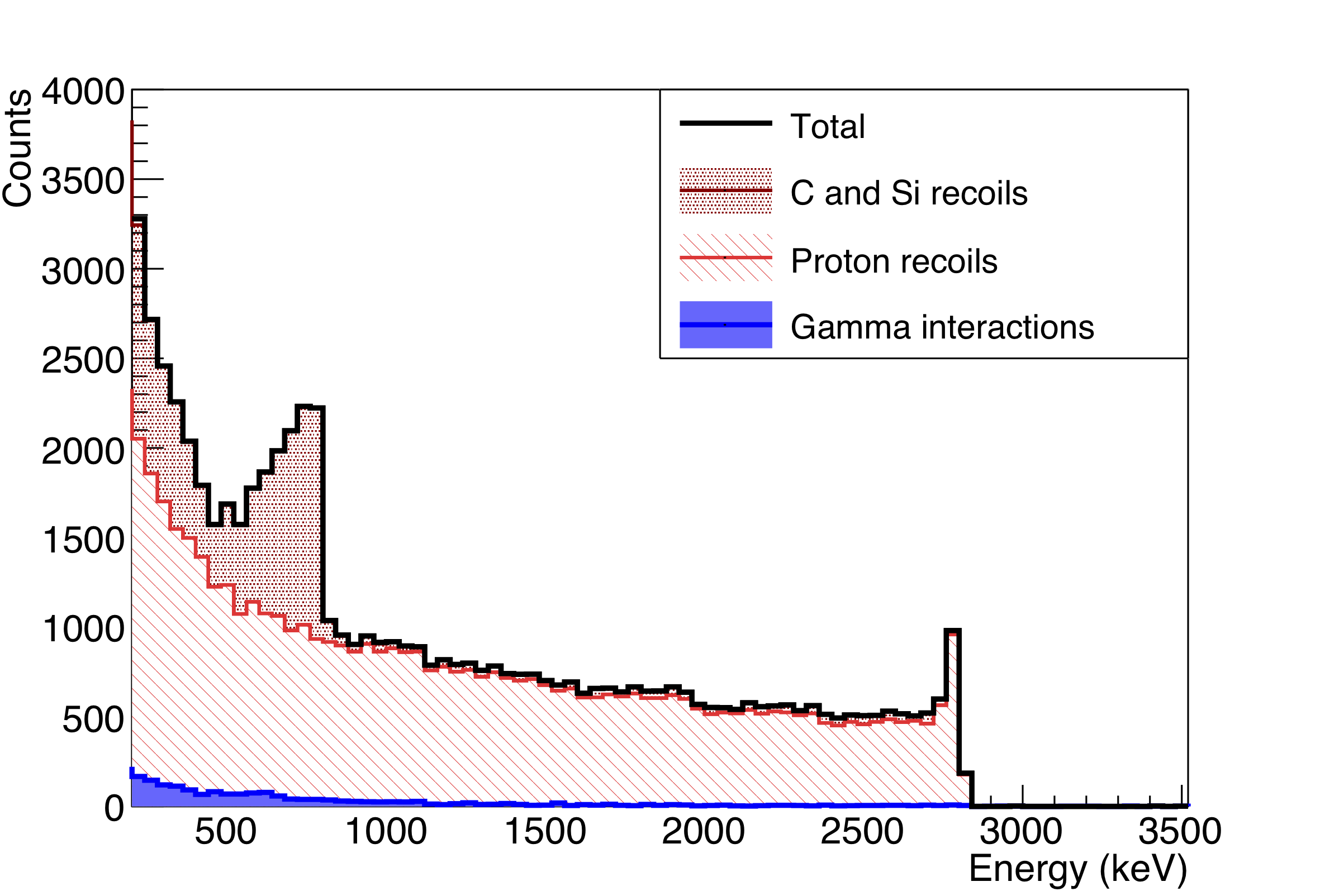}
    \caption{Simulated energy deposition from a single readout channel in the TPC. The total energy spectrum is shown by the black line. Gamma ray interactions (which are a background to the measurement) are shown in blue, while nuclear recoils are separated into two categories: proton recoils (red single hatch) and carbon/silicon recoils (red checkered).}
    \label{fig:MC}
\end{figure}

The simulated energy spectrum for a single channel is shown in Figure~\ref{fig:MC}. The simulation confirms that the gamma-ray background from neutron capture in the BPE and detector materials represents is subdominant in the nominal experimental configuration (although the background is estimated directly from the data, as explained in Sec.~\ref{subsec:reconstruction_and_selection}). The proton recoil energy distribution at high energies can be approximated by a step distribution with a small peak near the endpoint, and has a rising tail at low energies due to neutrons losing energy in passive materials before scattering in the active volume. The spectrum also shows a feature at $\sim$700~keV from neutron-induced carbon recoils. 

\subsection{Event reconstruction and selection}
\label{subsec:reconstruction_and_selection}

\begin{figure}[h!]
    \centering
    \includegraphics[width=0.85\columnwidth]{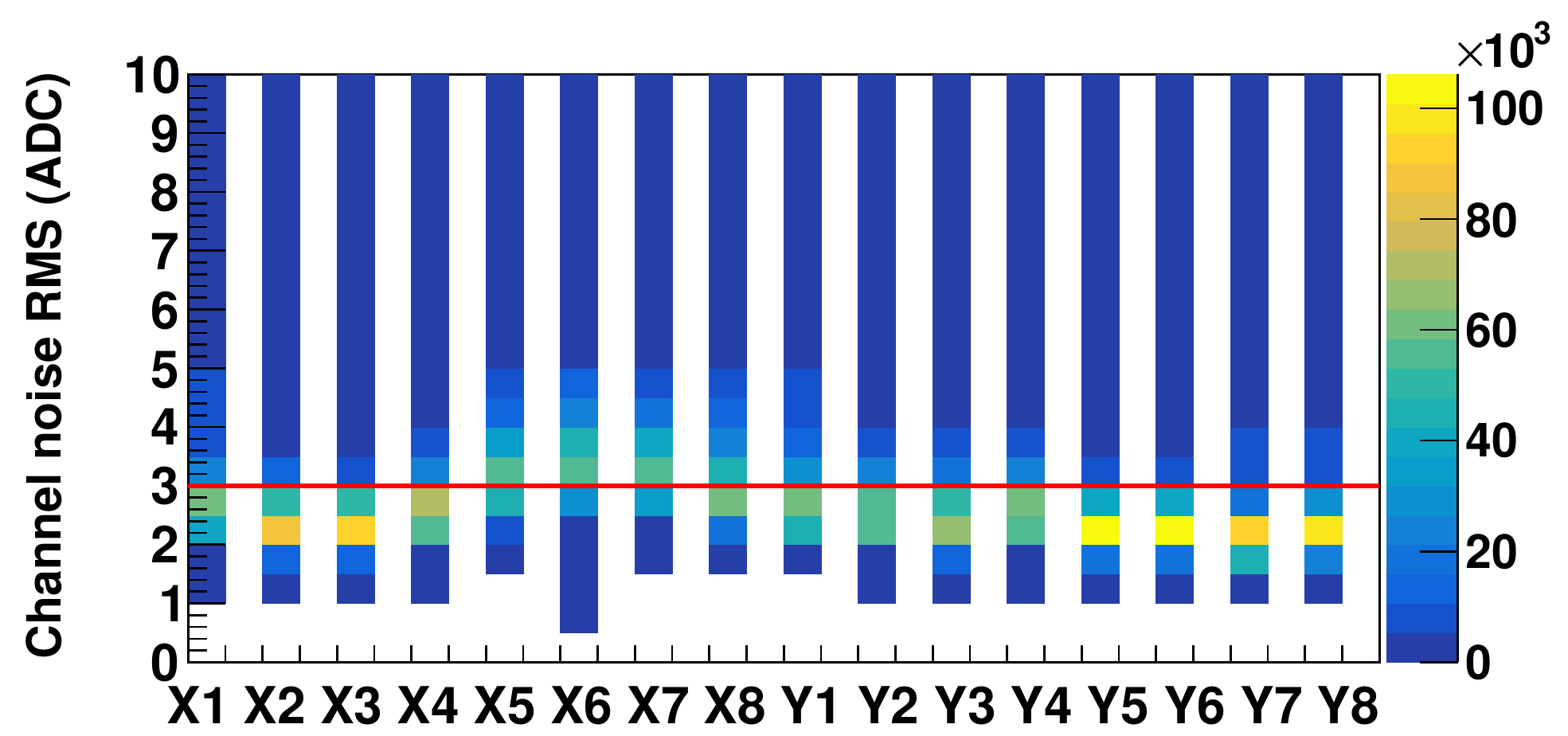}
     \includegraphics[width=0.85\columnwidth]{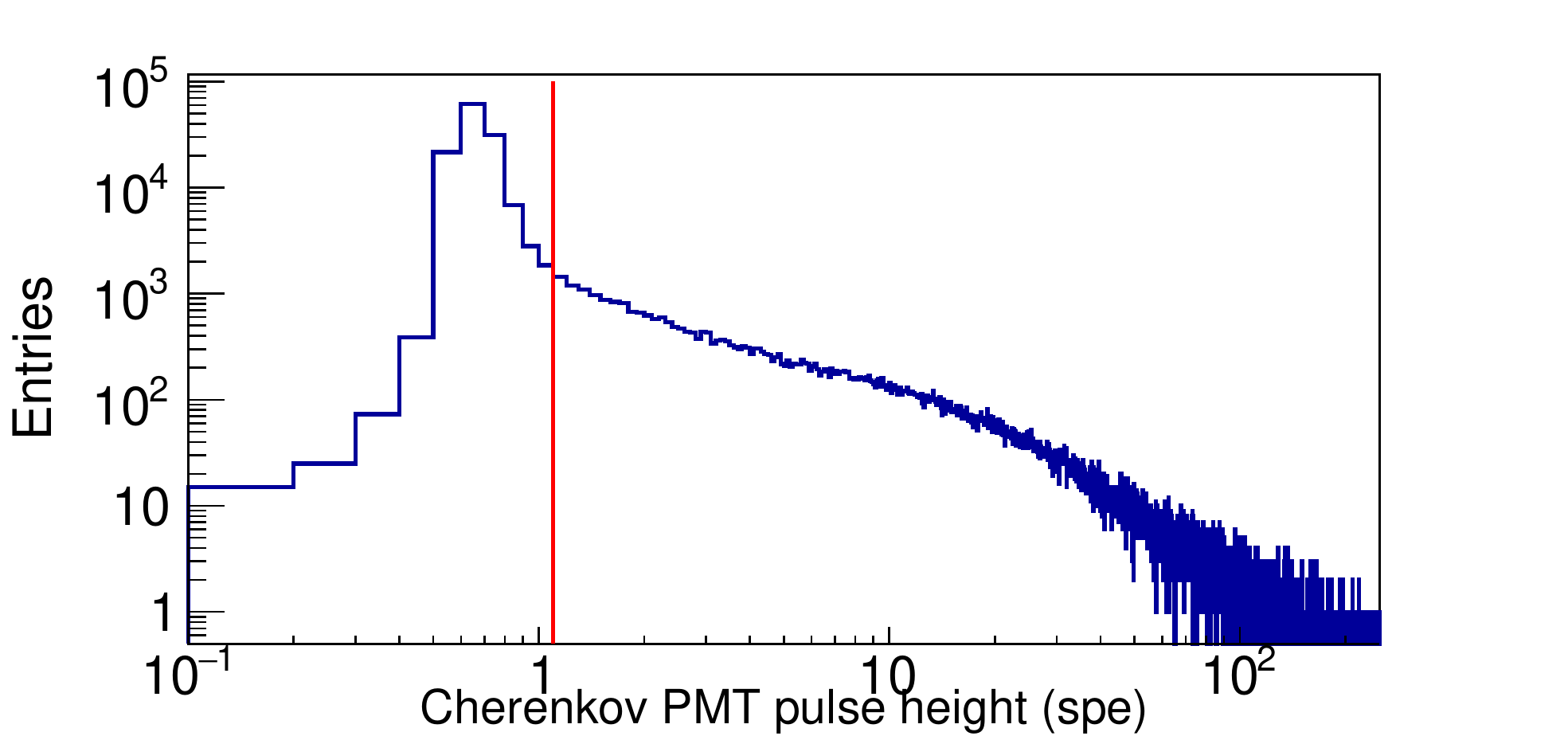}
    \caption{Selection cuts. Top: charge channel noise RMS and the ``hit" finding threshold in red.  Bottom: distribution of the Cherenkov PMT pulse height in number of single photon-electrons (spe).  The analysis requires this signal to be below the red line. }
    \label{fig:cuts}
\end{figure}
Triggers with the best noise conditions are selected by requiring that the RMS, calculated using 1000 samples from the pre-trigger window, not exceed 3 ADC counts. Hits are then identified at each of channels X2, X3, X4 when the corresponding waveforms exceed 3 times the RMS noise. The distribution of the RMS noise for each channel and the 3 ADC counts requirement are illustrated in Fig.~\ref{fig:cuts} (top).  Energy deposits are computed as the amplitudes of the steps in the signal from the charge integration on the X wires. This is achieved by averaging 500 samples after a trigger (between 33.6~$\mu$s and 41.6~$\mu$s) and subtracting the baseline, which is measured by averaging 1000 samples between 4.8~$\mu$s and 20.8~$\mu$s.  Since the reset time constant of the amplifiers is $\tau_{res} \simeq 500$~$\mu$s~\cite{Jewell_2018}, 500 samples correspond to $1.6\% \times \tau_{res}$, making the averaging a good estimator of the amplitude.  

Additional cuts are applied to select single-scatter nuclear recoil events and reject any residual backgrounds from gamma rays and cosmic ray muons. First, the number of each of X and Y wire hits is required to be 1. This removes muons which pass through the detector and create long tracks, as well as any particles which multiple-scatter within the active volume, such as Compton scattering gammas. Second, we remove events in which the signal in the Cherenkov PMT is above 1.1 single photon-electrons (spe) which is just above 3$\sigma$ of the noise in the PMT (see Fig.~\ref{fig:cuts}, bottom). This cut further suppresses cosmic rays and electrons with energy above 290~keV (typically produced in Compton scattering). Finally, we require that the time difference between the wire signal and the start of the beam gate to be between 500 and 800 samples (i.e. 8 -- 12.8~$\mu$s in Fig.~\ref{fig:timing}, top) to ensure that events are in coincidence with the neutron pulse. 

The background contribution is estimated by shifting the event selection time window to the interval between 0 and 300 samples (i.e. from 0 -- 4.8~$\mu$s in Fig.~\ref{fig:timing}, top). 

\begin{figure}[ht]
    \centering
    \includegraphics[width=0.8\columnwidth]{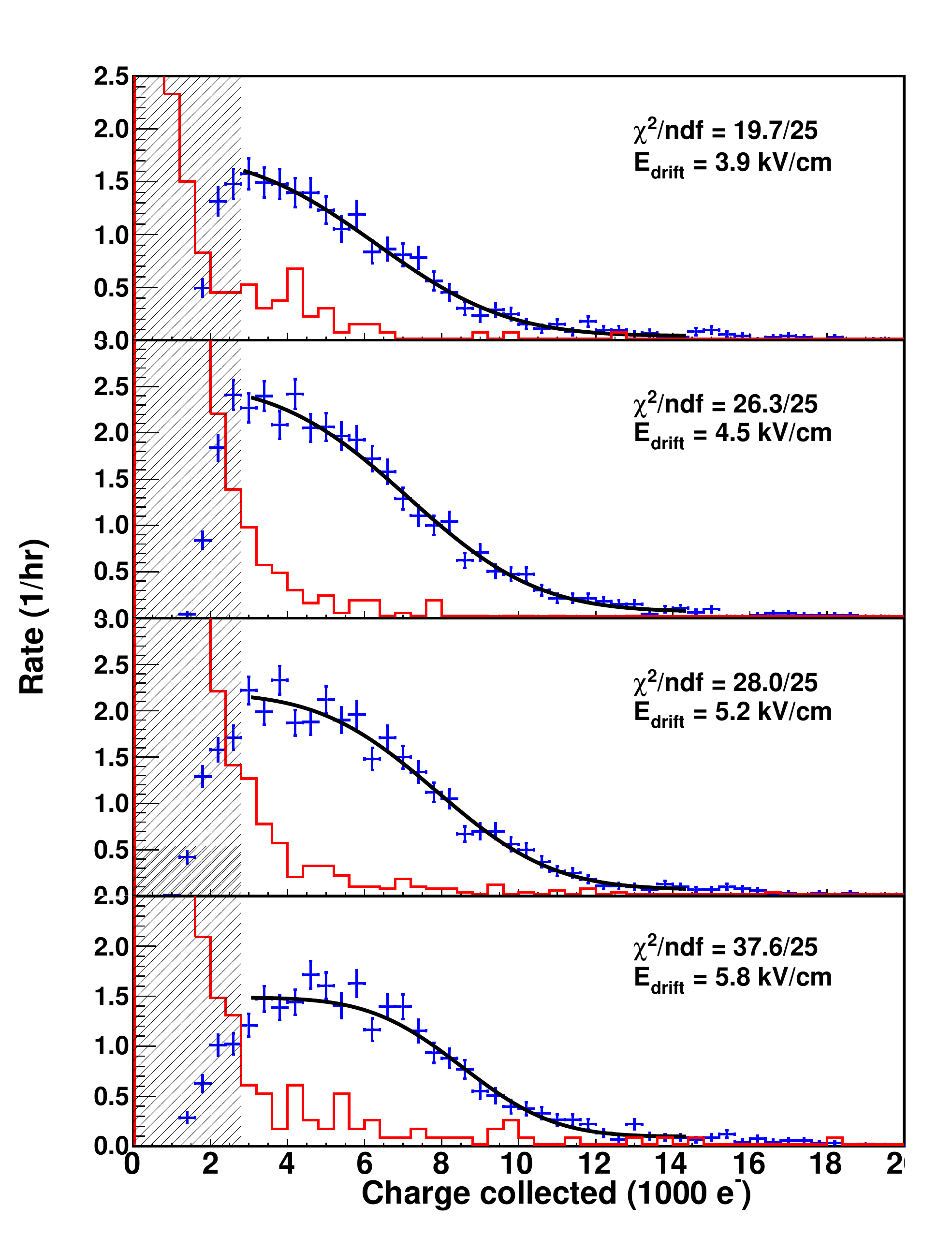}
    \caption{Energy deposition on channel X2 (blue crosses) after applying all selection cuts and the calibration obtained in~\cite{TMS_paper}. The four panels represent the different drift fields, as indicated. The grey areas represent the regions below the hardware threshold, and the noise spectra are shown in red.  The black curves are fits to steps convolved with Gaussuan functions.  The $\chi^2$ of the fits are also shown.}
    \label{fig:fit_vertical}
\end{figure}

\section{Results and discussions}

\begin{figure}[ht]
    \centering
    \includegraphics[width=0.8\columnwidth]{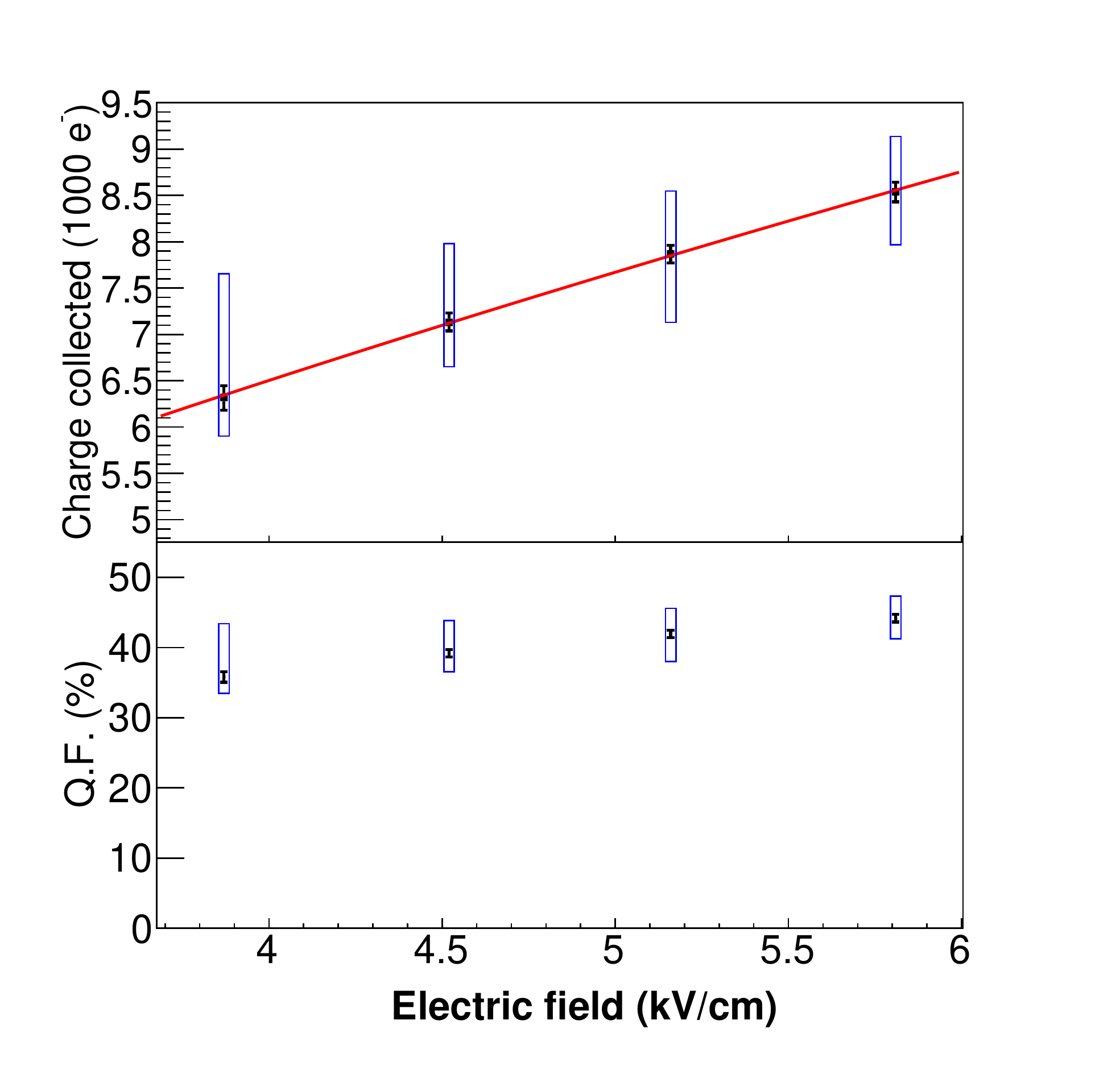}
    \caption{Top: charge yield at different drift fields together with the fit function using the Thomas-Imel model (red line). Bottom: proton recoil quenching factors. In all cases, systematic (statistical) errors are plotted as blue boxes (black bars). 
        }
    \label{fig:chargeyield}
\end{figure}

Results are extracted from the charge collection channel with the lowest electronics noise, X2. The energy distribution from neutron recoil events is obtained by subtracting from the spectrum of in-time events that of off-time ones, which is obtained from the background region in Figure~\ref{fig:timing}.  This results in the distributions shown in Figure~\ref{fig:chargeyield}, for the different applied electric fields.  
The absolute calibration in terms of charge is derived from the previous measurements reported in Ref.~\cite{TMS_paper} using the same detector and electronics. 
The distributions show the expected edge structure from proton recoil events, with a cutoff at $\sim$3000 electrons due to the hardware threshold.  Carbon and silicon recoils are expected to be below threshold. A noise spectrum, obtained by averaging 500 samples (different from the 1000 samples used to obtain the baseline) in the pre-trigger window, is also plotted in red, showing a clear difference in shape between the spectra for the same total number of events. The recoil spectrum above threshold is fitted to a step convolved with a Gaussian function. This parameterization approximates the proton recoil distribution shown in Figure~\ref{fig:MC} and models the fluctuations in the observed charge, which are due to a combination of physical processes in the TMS (e.g. recombination fluctuations) and the resolution introduced by the electronics and energy reconstruction. The fitting function has the analytical form:

\begin{center}
$f(E)=\frac{A}{2}(1-erf(\frac{(E-E_{E})}{\sqrt{2}\sigma}))$,
\end{center}
where $E$ is the energy deposited and the free parameters are $A$, the step amplitude, $E_{E}$, the edge location, and $\sigma$, the Gaussian energy resolution.

\begin{table*}[h!!!!]
    \centering
    \begin{tabular}{c|c|c|c|c|c|c|c|c}
        Electric field  & $E_E$ & $\sigma$ & $\sigma/E_E$ & $E_E$ Stat. & $E_E$ Sys. & Q.F. & Q.F. Stat. & Q.F. Sys. \\
        (kV/cm) & ($10^3$~e$^-$) &  ($10^3$~e$^-$) & (\%) & ($10^3$~e$^-$) & ($10^3$~e$^-$) & (\%) & (\%) & (\%) \\ 
        \hline
        \hline 
        & & & & & & & \\[-1pt]
        3.9 & 6.32 & 2.67 & 42.2 & $^{+0.13}_{-0.14}$ & $^{+1.33}_{-0.42}$ & 35.8 & $^{+0.73}_{-0.78}$  & $^{+7.6}_{-2.4}$ \\ [5pt]
        4.5 & 7.13 & 2.65 & 37.2 & $^{+0.10}_{-0.09}$ & $^{+0.85}_{-0.48}$ & 39.2 & $^{+0.55}_{-0.50}$ & $^{+4.7}_{-2.6}$ \\ [5pt]
        5.2 & 7.87 & 2.39 & 30.3 & $^{+0.09}_{-0.10}$ & $^{+0.68}_{-0.74}$ & 42.0 & $^{+0.48}_{-0.53}$ & $^{+3.6}_{-3.9}$\\ [5pt]
        5.8 & 8.54 & 1.89 & 22.1  & $^{+0.10}_{-0.11}$ & $^{+0.60}_{-0.57}$ & 44.2 & $^{+0.53}_{-0.57}$  &  $^{+3.1}_{-3.0}$ \\ [5pt]
    \end{tabular}
    \caption{Fit results obtained for the data shown in Fig.~\ref{fig:fit_vertical} and Fig.~\ref{fig:chargeyield}, along with statistical and systematic uncertainties for the edge location $E_E$.  The quenching factors (Q.F.) are also shown, along with their uncertainties.}
    \label{tab:fitvalues}
\end{table*}

The fitted values for $E_E$ and $\sigma$ at each electric field are summarized in Table~\ref{tab:fitvalues}, together with the associated uncertainties for $E_E$. Four sources of systematic errors are considered: the choice of fitting range and binning, the calibration of the readout system, and the correlation between energy resolution and the edge position from the fit. The latter is potentially important, because of the absence of a very specific feature in the spectrum.  The fit range is varied by up to 800~$e^-$ in either direction, resulting in a change in the fitted endpoint position of up to 3.5\%. Changes to the binning scheme result in changes of up to 6.7\%. The energy calibration in Ref.~\cite{TMS_paper}, used to scale the measured signals to units of electrons, provides another 6.6\% uncertainty. The systematic effect deriving from the correlation between energy resolution ($\sigma$) and the edge position ($E_E$) is estimated by fixing the value of $\sigma$ over a range of values and repeating the fit, resulting in an uncertainty in $E_E$ of 5\% (20\%) at the highest (lowest) electric field.  The final reported systematic uncertainty is the quadrature sum all systematics, added independently for the positive and negative directions.

The charge yield for 2.8~MeV neutrons as a function of the electric field is shown in Fig.~\ref{fig:chargeyield} (top).   From a comparison with the recently reported charge yield for $\gamma$s~\cite{Farrad_che_2018}, the proton recoil quenching factor (Q.F.) is obtained for each electric field, as shown in Fig.~\ref{fig:chargeyield} (bottom) and listed in Table~\ref{tab:fitvalues}.  These field-dependent quenching factors range from 36\% to 44\% and are comparable to those ($\sim 30\%$) measured for plastic scintillators with a similar chemical composition~\cite{WELDON2020163192, Awe_2021}, presumably implying that the quenching is universally dominated by the migration of energy into channels other than ionization or scintillation.

The electron-ion recombination probability in cryogenic noble liquids (e.g. LAr and LXe) is often described using the Thomas-Imel box model, which gives the number of ionized electrons after recombination, $N_{e}$, as:
\begin{equation}
\label{eq:thomasimel}
N_{e}=\frac{N_i}{\xi}ln(1+\xi), \hspace{1cm} \xi = \frac{N_iC}{E},
\end{equation}
where $N_{i}$ is the initial number of electron/ion pairs prior to recombination, and $C$ is a constant. We fit this expression to our measurements allowing $N_{i}$ and $C$ to vary, finding $N{_i}=78676\pm45720$ and $C=0.24\pm0.05$. We note that the initial number of ions $N_i$ is poorly constrained, which we attribute to the fact that the range of drift fields used in this work is far below the saturation field in TMS~\cite{engler_1986}. However, the fitted value of $C$ is well-constrained. 

In the Thomas-Imel derivation, the constant $C$ has the form of $\alpha/4a^2\mu$, with $\alpha$ being the recombination coefficient, $\mu$ being the mobility, and $a$ being the dimension of the box where the initial charge is uniformly distributed in the model. Using the Langevin parametrization of the recombination coefficient, $\alpha$ is given by
$\alpha= e\mu/\epsilon$,
where $\epsilon$ is the dielectric constant of the liquid. Taking the fitted value for $C$, one can obtain the box dimension parameter $a$, which quantifies an effective length scale across which electron/ion recombination can occur. We find $a\sim10~\mu$m, which can be compared with the $\mathcal{O}(100\;\mathrm{nm})$ reported mean thermalization length of electrons produced by heavily ionising particles~\cite{ENGLER1992479}. The fitted box dimension parameter is significantly larger than the mean thermalization length, in agreement with the assumption of non-geminate electron-ion recombination in the Thomas-Imel model.

In noble liquid applications, the field dependence of the Thomas-Imel model is often modified by replacing the $1/E$ dependence in Equation~\ref{eq:thomasimel} with $1/E^b$, where $b$ is an additional free parameter which generally fits to a number significantly less than one~\cite{prl_Joshi,Dahl:2009nta,NEST_NR_model}. Making this modification and refitting to our data with $b$ floating gives $N{_i}=26810\pm14700$, C=$0.21\pm0.02$, b=$1.2\pm0.3$. While such a model is severely underconstrained with only four data points, we note that the fitted values of $N_i$ and $C$ are consistent with those obtained above, and the fitted value of $b$ is consistent with the $1/E$ field dependence derived by Thomas and Imel.

\section{Conclusion}
In this paper, we describe the detection of proton recoils from 2.8 MeV neutrons in a TPC filled with TMS. Using the endpoint of the measured ionization signals, we perform the first measurement of charge yield for low energy recoils in TMS.   By comparing with previous measurements using fast electrons, we extract the quenching factor as a function of the electric field in the TPC. We find that the Thomas-Imel box model provides a good description of the ionization yield as a function of the electric field and, from the measured parameters, deduce that columnar recombination plays an important role in the conditions of these measurements. The quenching factor obtained is a useful input for evaluating possible applications of warm liquid ionization chambers, including detectors of fast neutrons or of other low energy phenomena. 

\acknowledgments
We thank J.~Lin and D.~McKinsey (UC Berkeley) for loaning the neutron generator used in this work, and for their help with its operation. We also thank E.~Angelico and C.~Hardy for their comments on this manuscript. S.~W. and B.~L. acknowledge the partial support of the Karl A. Van Bibber Postdoctoral Fellowship from Stanford physics department. This work was supported by seed funds of Stanford University. 
 
%\bibliographystyle{model1-names}
%\bibliographystyle{apsrev4-2}
%\printbibliography[maxnames=2]
%\bibliographystyle{plain}
%\bibliographystyle{apsrev}
%\bibliography{bibfile_short.bib}

\end{document}